\documentclass[acmsmall,screen,nonacm]{acmart}

\AtBeginDocument{%
  }

\usepackage{booktabs}
\usepackage{multirow}
\usepackage{xspace}
\usepackage{xcolor}
\usepackage{caption}
\usepackage{algorithm}
\usepackage{algpseudocode}
\usepackage{tcolorbox}
\usepackage{wrapfig}
\usepackage{enumitem}

\newcommand{\system}{\textsc{DesignLore}\xspace}
\newcommand{\benchmark}{\textsc{DesignReviewBench}\xspace}

\newcommand{\boxmargin}{1mm}
\newtcolorbox{mybox}{
    colback=gray!15!white,
    arc = 0pt, outer arc = 0pt,
    boxsep=0pt, left = 3pt, right = 0pt, top = 0pt, bottom = 0pt, 
    leftrule=3pt, bottomrule=0pt,toprule=0pt, rightrule=0pt,
    left = \boxmargin, right = \boxmargin, top = \boxmargin, bottom = \boxmargin
}
\newcommand{\findingbox}[2]{%
\begin{center}
    \begin{mybox}{\textbf{#1: } #2}
    \end{mybox}
\end{center}
}
\newcommand{\revise}[1]{\textcolor{black}{#1}}

\newcommand{\rqsummary}[2]{%
\begin{center}
    \begin{mybox}{\textbf{#1 Summary: } #2}
    \end{mybox}
\end{center}
}

\newcommand{\distance}{3pt}
\begin{document}

\title[Benchmarking and Enhancing Design-Aware Code Review in the Linux Kernel]{Benchmarking and Enhancing Design-Aware Code Review\\ in the Linux Kernel}

\author{Zhenhao Zhou}
\affiliation{%
  \institution{Fudan University}
  \city{Shanghai}
  \country{China}}
\email{zhzhou24@m.fudan.edu.cn}

\author{Kai Yu}
\affiliation{%
  \institution{Fudan University}
  \city{Shanghai}
  \country{China}}
\email{23262010042@m.fudan.edu.cn}

\author{Ying Wang}
\affiliation{%
  \institution{Fudan University}
  \city{Shanghai}
  \country{China}}
\email{yingwang25@m.fudan.edu.cn}

\author{Zhiqiang Yuan}
\affiliation{%
  \institution{Fudan University}
  \city{Shanghai}
  \country{China}}
\email{zhiqiangyuan23@m.fudan.edu.cn}

\author{Junhao Zeng}
\affiliation{%
  \institution{Fudan University}
  \city{Shanghai}
  \country{China}}
\email{jhzeng25@m.fudan.edu.cn}

\author{Yijian Wu}
\affiliation{%
  \institution{Fudan University}
  \city{Shanghai}
  \country{China}}
\email{wuyijian@fudan.edu.cn}

\author{Chong Wang}
\affiliation{%
  \institution{Nanyang Technological University}
  \city{Singapore}
  \country{Singapore}}
\email{chong.wang@ntu.edu.sg}

\author{Xin Peng}
\authornote{Corresponding author.}
\affiliation{%
  \institution{Fudan University}
  \city{Shanghai}
  \country{China}}
\email{pengxin@fudan.edu.cn}

\begin{abstract}
Code review often requires reasoning beyond local correctness. In mature
software systems, reviewers must recognize design trade-offs introduced by
a proposed change, compare alternative solutions, and reason about their
compatibility with project-specific constraints. We refer to this problem
as \emph{design-aware code review}. However, existing automated review
approaches primarily focus on defects and correctness, while the design
knowledge accumulated in historical review discussions remains difficult to
reuse.

To address these challenges, we present \system, a knowledge-augmented
approach that extracts design decisions from historical review discussions,
organizes them into an adaptive design taxonomy, and reuses relevant
historical knowledge to guide code review. We further construct \benchmark,
a benchmark derived from real patch-review histories for systematically
evaluating design-aware code review. Our evaluation shows that design-aware
review remains challenging for current coding agents, with the strongest
baseline achieving an F1 of only 0.213 on design trade-off detection.
In comparison, \system achieves an F1 of 0.259, a 21.6\% relative improvement,
and consistently  delivers gains across different agent systems and backbone. 
These results highlight the importance of design-aware review and demonstrate 
the potential of recovering and reusing historical design knowledge to support it.
\end{abstract}



\authorsaddresses{}

\maketitle

\section{Introduction}
\label{sec:introduction}

Code review is a central quality-assurance activity in software development, particularly in long-lived systems such as the Linux kernel, where reviewers examine proposed changes before integration to identify and resolve potential problems~\cite{bacchelli2013expectations}. With recent advances in large language models (LLMs) and LLM-based agents~\cite{DBLP:journals/corr/abs-2406-00515,DBLP:journals/chinaf/ZhangFXZYSYC26,DBLP:journals/corr/abs-2409-02977}, a growing number of automated code-review approaches and benchmarks have been proposed~\cite{li2022codereviewer,lin2025codereviewqa,zeng2026swrbench,li2026opencodereview,DBLP:journals/corr/abs-2603-23448,DBLP:journals/corr/abs-2603-24825}. These efforts, including the recent agent-based Sashiko~\cite{sashiko2026}, primarily focus on functional correctness or defects of the code changes.

\begin{figure*}[t]
  \centering
  \includegraphics[width=\textwidth]{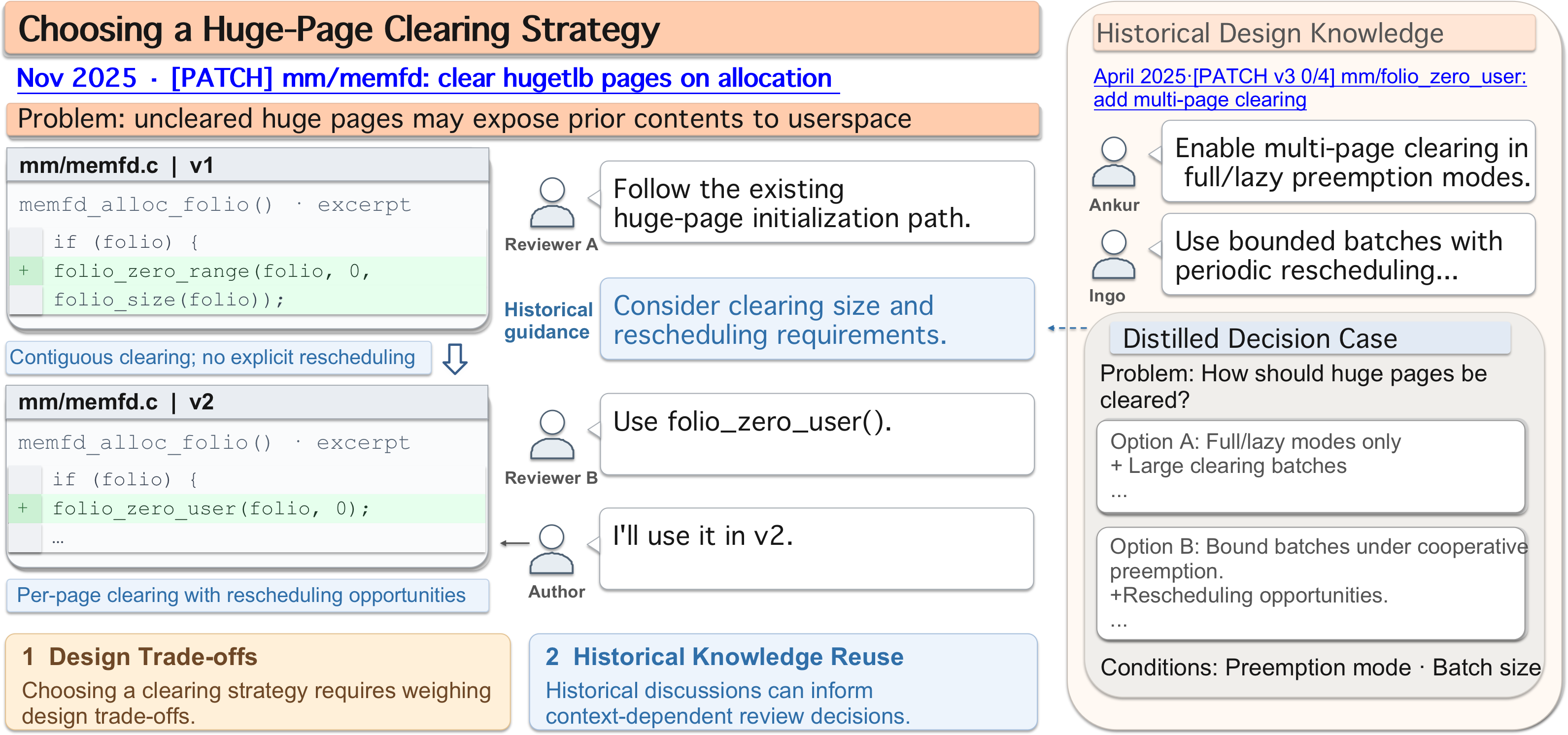}
  \caption{A motivating example from the Linux kernel \texttt{mm} subsystem.}
  \label{fig:motivation}
\end{figure*}

In practice, effective review often extends beyond detecting local defects or checking functional correctness to evaluating \textbf{\emph{design trade-offs}}. A proposed change reflects one design choice, but alternative solutions may also satisfy the same functional requirements with different advantages. Reviewers therefore need to understand the trade-offs among viable alternatives and determine which choice better fits the project's design principles and long-term needs. This consideration is particularly important in Linux kernel development, where development guidance explicitly asks contributors to consider how code will be maintained years after it is merged~\cite{linuxkernel2023process}. For example, when clearing huge pages in the \texttt{memfd} patch shown in Figure~\ref{fig:motivation}, reviewers suggested replacing \texttt{folio\_zero\_range()} with \texttt{folio\_zero\_user()}. Both are functionally correct but make different trade-offs: the former favors throughput, while the latter favors scheduling responsiveness. The choice depends on the preemption configuration rather than functional correctness alone.

We refer to this problem as \textbf{\emph{design-aware code review}}, which goes beyond local correctness checking to identify and reason about the design trade-offs underlying code changes. Such reasoning may draw on information recorded in historical review discussions~\cite{elzanaty2018design,bi2021architecture,DBLP:conf/acl/ZhouHHWWWPL26}. In the Linux kernel, these discussions take place on public mailing lists and capture design considerations at multiple levels of granularity, including the alternatives considered, their advantages and disadvantages, and the reasoning behind the final decisions. They may therefore offer useful context when similar design problems arise in future patches. For example, as shown in Figure~\ref{fig:motivation}, an April 2025 discussion of a different patch series considered the trade-off between bulk clearing and scheduling responsiveness, with a reviewer proposing bounded clearing batches with periodic rescheduling. The discussion thus provides relevant context for the later choice in the \texttt{memfd} patch.

However, whether this opportunity matters in practice remains unclear: \emph{how frequently code review entails design trade-offs and how often similar design problems recur in ways that make historical decisions relevant.} To answer these questions, we conduct a formative study on the memory-management (\texttt{mm}) subsystem of the Linux kernel, whose public review process provides rich historical discussions and traceable patch revisions. Analyzing design decisions across many review threads through manual inspection alone is impractical. Relevant reasoning is often fragmented across multiple replies, interleaved with implementation details and procedural discussions, and expressed at different levels of abstraction. We therefore propose an automated pipeline that reconstructs each design decision discussed in a thread, together with the alternatives considered and their trade-offs, into a structured \textbf{\emph{decision case}}. Applying it to 600 threads sampled between 2021 and 2026 yields 1,774 decision cases, whose quality we validate through human evaluation.

Based on the mined cases, the study confirms that design trade-offs are common in practice: \textbf{60.29\%} of the sampled threads involve at least one design trade-off, and \textbf{33.91\%} contain a problem with a relevant historical precedent. It also shows that exploiting such precedents is non-trivial: the reasoning behind a decision must first be recovered from fragmented discussions, a relevant precedent often concerns different code and terminology than the new patch, and a historical resolution applies only if its conditions still hold in the current code. The challenge is therefore not the absence of historical design knowledge, but how to recover, organize, and reuse it effectively.

Guided by the findings, we present \system, a knowledge-augmented approach for design-aware code review. \system first applies the extraction pipeline to review history to build a knowledge base of decision cases, then organizes them into an adaptive, project-specific design taxonomy based on the code patterns that trigger each case, enabling precedents to be retrieved from a new patch's code rather than surface text. At review time, \system uses the taxonomy and relevant historical decisions to guide the analysis of a target patch, helping the agent identify and validate design trade-offs against the current code and project context.

We construct \benchmark from real patch-review histories of the Linux kernel, comprising 72 patch series and 79 reference design trade-offs. Each benchmark instance pairs a proposed patch series with the codebase snapshot available at review time and captures the design trade-offs revealed by subsequent human review. We compare \system with state-of-the-art general-purpose coding agents, specialized code-review systems, and retrieval-augmented agents that access raw historical review emails. The results show that design-aware code review remains challenging for current agents: the strongest baseline, Sashiko~\cite{sashiko2026}, achieves an F1 of 0.213. In contrast, \system achieves the best trade-off detection performance, with an F1 of 0.259, a 21.6\% relative improvement at roughly half of Sashiko's review cost. The improvement persists across three agent systems and four backbone models. The cumulative ablation shows incremental F1 improvements, with the largest increase observed when adding the planning and validation workflow.

In summary, this paper makes the following contributions:

\begin{itemize}[leftmargin=12pt, topsep=3pt, itemsep=2pt, parsep=0pt, partopsep=0pt]
  \item \textbf{An empirical characterization of design-aware code review.}
  We analyze historical Linux kernel reviews to characterize the prevalence and recurrence of design trade-offs and the challenges of reusing historical design knowledge.

  \item \textbf{A knowledge-augmented agentic code review approach.}
  We present \system, which extracts design knowledge from historical reviews, organizes it into an adaptive design taxonomy, and reuses relevant decisions to guide new patch reviews.

  \item \textbf{A benchmark for design-aware code review.}
  We construct \benchmark from real historical patch-review discussions to evaluate whether coding agents can identify design trade-offs exposed by proposed changes.

  \item \textbf{A comprehensive empirical evaluation.}
  We evaluate \system against general-purpose, review-specialized, and retrieval-augmented coding agents, demonstrating its effectiveness across diverse agents and backbone models.
\end{itemize}

\section{Formative Study}
\label{sec:formative-study}

We conduct a formative study to understand what design-aware code review entails in practice and whether historical review discussions can support it. We aim to investigate three questions:
\begin{itemize}[leftmargin=12pt, topsep=3pt, itemsep=2pt, parsep=0pt, partopsep=0pt]
  \item \textbf{FQ1 (Prevalence):} How frequently do review decisions in
  Linux kernel patch review involve design trade-offs?
  \item \textbf{FQ2 (Recurrence):} How often have similar underlying design
  decisions already appeared in earlier, independent patch series?
  \item \textbf{FQ3 (Implications):} What challenges arise in reusing
  historical review discussions for code review, and what do they imply for
  automated support?
\end{itemize}
\noindent FQ1 reflects the practical need for reasoning beyond local
correctness, whereas FQ2 indicates whether historical review discussions can
provide reusable precedents for future reviews. FQ3 is answered by
reflecting on the difficulties encountered when constructing decision cases and
identifying precedents for FQ1 and FQ2.

\subsection{Motivation and Definition of Decision Cases}\label{sec:decision-case-definition}

Raw review threads do not provide a consistent unit of analysis:
one thread may contain several independent design questions,
while reasoning about one question may span multiple reviewers and replies
and patch revisions. We therefore reconstruct discussions into
\emph{decision cases}, analyze trade-offs and recurrence at
the case level, and aggregate the results to the thread level.

We define a \emph{decision case} as a historical record of a design decision and the deliberation leading to it, including the alternatives considered and their trade-offs. Each case comprises:

\begin{itemize}[leftmargin=12pt, topsep=3pt, itemsep=2pt, parsep=0pt, partopsep=0pt]
\item \textbf{Problem:} The technical question or design tension motivating the decision.
\item \textbf{Code-pattern trigger:} The code context or recurring pattern giving rise to the problem.
\item \textbf{Options:} The alternatives considered, each with a \textbf{Description} of the proposed action or position, \textbf{Conditions} under which it applies, \textbf{Pros} describing its expected benefits, and \textbf{Cons} describing its limitations, risks, or trade-offs.
\item \textbf{Decision:} The option ultimately agreed upon by the participants.
\item \textbf{Rationale:} The justification for the decision.
\end{itemize}

\subsection{Decision Case Mining}\label{sec:decision-extraction}
\begin{figure*}[t]
  \centering
  \includegraphics[width=\textwidth]{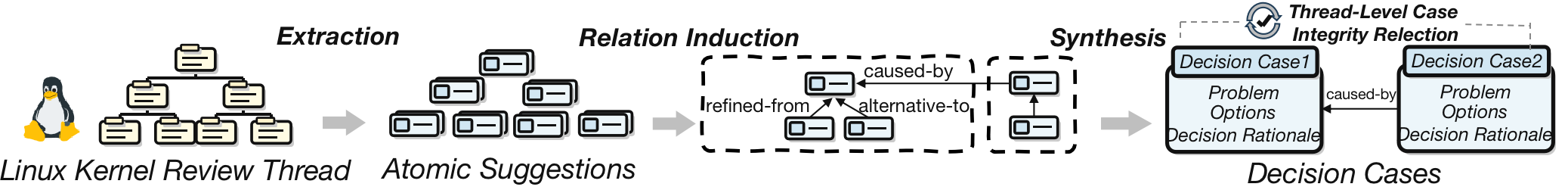}
  \caption{Overview of Decision Case Mining.}
  \Description{
  Offline decision-knowledge extraction transforms historical review threads
  into source-linked decision cases.}
  \label{fig:extract-pipeline}
\end{figure*}

We propose a four-step mining pipeline that progressively transforms fine-grained atomic suggestions into coherent decision cases, as illustrated in Figure~\ref{fig:extract-pipeline}.

\subsubsection{Atomic Suggestion Extraction}
\label{sec:atomic-extraction}

We first use an LLM to extract source-grounded \emph{atomic suggestions}
from review threads, where each suggestion addresses a single independently
decidable technical concern. A suggestion may defend or refine the current
design, object to an existing proposal, introduce an alternative, or raise a
substantive design question. Each suggestion records the \emph{Problem} and
\emph{Code-pattern trigger} fields of a decision case, together with the
\emph{Suggestion}, \emph{Arguments} for or against it, and applicable
\emph{Conditions}.

Review discussions are tree structured through email replies, which does not
directly align with the sequential context expected by autoregressive LLMs.
We therefore organize each thread into bounded local contexts before
extraction. Formally, let a review thread be represented as a rooted reply
tree $T=(V,E)$, where each node $v\in V$ denotes an email and each directed
edge $(p(v),v)\in E$ denotes that $v$ directly replies to its parent $p(v)$.
For each target email $v$, a local window of size $s$ is constructed from the
target email and its nearest ancestors along the reply path:
$W(v)=\langle v, p(v), p^{2}(v), \ldots, p^{s-1}(v)\rangle$,
where the sequence is truncated when the root of the thread is reached. In
practice, $s=2$ is used to preserve local conversational context while keeping
the extraction scope compact.

Given a window $W(v)$, zero or more atomic suggestions are extracted from the
target email using the LLM. Only text newly written by the target author is
treated as extractable evidence; the parent email and quoted text within the
target email are provided as read-only context for resolving references and
interpreting the discussion. The LLM is instructed to focus on substantive
technical content and ignore purely procedural remarks, such as
acknowledgements and requests to resend a patch. Each extracted suggestion
retains provenance to its source email and supporting text, enabling
subsequent relation construction and independent verification.

\begin{table}[t]
\caption{Logical relations between atomic suggestions.}
\label{tab}
\centering
\small
\begin{tabular}{p{0.20\linewidth}p{0.70\linewidth}}
\toprule
Relation & Semantics \\

\midrule
\texttt{refined-from} &
Addresses the same technical question and follows the same solution
direction, while adding detail, evidence, clarification, or an
applicability condition. \\
\texttt{alternative-to} &
  Addresses the same technical question but proposes or defends a
  competing solution direction, including retaining the existing design. \\
\texttt{caused-by} &
  Raises a distinct technical question that is exposed by the preceding
  discussion and requires an independent resolution. \\
\bottomrule
\end{tabular}
\end{table}

\subsubsection{Reply-Constrained Logical Relation Induction}
\label{sec:relation-induction}

The discussion process is recovered by identifying how suggestions relate to
one another. The email reply structure is used as a strong prior to avoid
comparing every pair of suggestions, restricting the candidate predecessors
of each suggestion to those permitted by the reply topology. An LLM then
identifies the most relevant parent suggestion among these candidates and
classifies their logical relation into one of three types:
\texttt{refined-from}, \texttt{alternative-to}, or \texttt{caused-by}.
Table~\ref{tab} summarizes their semantics. Applying these relations across
the thread produces a typed directed tree over atomic suggestions that
captures both the reply structure and the inferred design-level relations.

\subsubsection{Decision Case Synthesis}
\label{sec:case-synthesis}

\begin{wrapfigure}{r}{0.52\textwidth}
\vspace{-10pt}
\begin{minipage}{0.52\textwidth}
\begin{algorithm}[H]
\caption{Relation-guided suggestion clustering}
\label{alg:suggestion-clustering}
\footnotesize
\begin{algorithmic}[1]
\Require Typed atomic-suggestion tree $T=(V,E)$
\Ensure Issue clusters $\mathcal{C}$ and causal edges $E_C$

\State create cluster $C_0$ for root $r$
\State assign $r$ to $C_0$
\State $\mathcal{C} \gets \{C_0\}$, $E_C \gets \emptyset$

\For{each edge $(u,v)$ in depth-first order}
\If{$\mathrm{rel}(u,v) \in
\{\texttt{refined-from},\texttt{alternative-to}\}$}
\State assign $v$ to the cluster of $u$
\ElsIf{$\mathrm{rel}(u,v)=\texttt{caused-by}$}
\State create a new cluster $C_v$ containing $v$
\State $\mathcal{C} \gets \mathcal{C} \cup \{C_v\}$
\State $E_C \gets E_C \cup
\{(\mathrm{cluster}(u),C_v)\}$
\EndIf
\EndFor
\State \Return $\mathcal{C}, E_C$
\end{algorithmic}
\end{algorithm}
\end{minipage}
\vspace{-5pt}
\end{wrapfigure}

As shown in Algorithm~\ref{alg:suggestion-clustering}, suggestions concerning
the same technical question are deterministically grouped by a depth-first
traversal of the typed tree. A child linked by \texttt{refined-from} or
\texttt{alternative-to} continues the same decision problem and joins its
parent's cluster, whereas a \texttt{caused-by} child opens a new cluster and
records a causal edge between the two clusters. The suggestions in each
cluster are then synthesized into a single decision case.

For each cluster, an LLM is given the constituent atomic suggestions together with their provenance and inferred relations, and is instructed to reconstruct the decision case following a fixed schema. The problem and code-pattern trigger are consolidated, and
the suggestions are processed chronologically to recover the alternatives
considered. Suggestions that duplicate, refine, or complement the same
solution are consolidated into one option, whereas suggestions advocating
different solution directions are retained as separate options. To preserve
the completeness of the decision space, the initial proposal embodied by the
reviewed patch is also recovered as an option. Arguments supporting an option
are consolidated into its \emph{Pros}, while objections and reported
drawbacks are consolidated into its \emph{Cons}. Applicability conditions
associated with the constituent suggestions are likewise propagated to the
corresponding options. The final decision is identified as the adopted option
based on explicit agreement or, when such agreement is absent, a subsequent
revision that clearly implements the option. If neither provides sufficient
evidence, the outcome remains unresolved. The rationale is reconstructed from
the arguments supporting the adopted option. Finally, cluster-level
\texttt{caused-by} relations are transferred to the corresponding decision
cases to construct the decision-case tree. All reconstructed fields retain
links to their supporting suggestions and source messages.

\subsubsection{Thread-Level Case Integrity Relection}
\label{sec:case-check}
An LLM is then given the reconstructed decision-case tree and is instructed to inspect the tree top-down for thread-level structural inconsistencies. 
The decision-case tree is traversed to check whether each case forms a coherent and independently resolvable decision problem, and whether the
relations between cases are consistent with the overall discussion. When an
inconsistency is detected, one of three structural repairs is applied. First,
\emph{merge} combines cases that were separated locally but in fact describe
different parts of the same decision problem. Second, \emph{reparent}
introduces or corrects a \texttt{caused-by} relation when two cases were placed as siblings although one technical question was actually triggered by the
other. Third, \emph{split} partitions an overly broad case when it conflates
multiple independently resolvable questions. Each repair preserves the
provenance of the affected fields and relations, resulting in a decision-case
tree that provides a thread-level consistent representation.

\subsection{Mined Decision Cases and Annotation}
\label{sec:annotation}
We randomly sample 100 eligible review threads from each year from January 2021
through August 2026, yielding 600 threads in total. To account for differences
in review volume across years, yearly estimates are weighted by the number of
eligible threads in each year. Applying the pipeline with GPT-5.6 Luna as the
underlying LLM to the 600 sampled threads yields 1,774 decision cases. Before
using these cases for FQ1 and FQ2, their quality is first assessed through a
human evaluation.

\textit{Manual Quality Assessment.}
Fifty decision cases are randomly sampled from the mined cases. Ten developers
with at least three years of programming experience and basic familiarity with
operating systems are recruited to independently assess the sampled cases.
None of the participants had prior involvement in this work. Each case is
rated on a four-point Likert scale (higher is better) along three dimensions:
\emph{correctness} (the content is faithfully supported by the original
discussion), \emph{completeness} (the information needed to understand the
problem and decision is captured), and \emph{conciseness} (irrelevant or
redundant information is avoided). Ratings of 1 or 2 require a brief
explanation.

\begin{wrapfigure}{r}{0.52\textwidth}
    \centering
    \includegraphics[width=\linewidth]{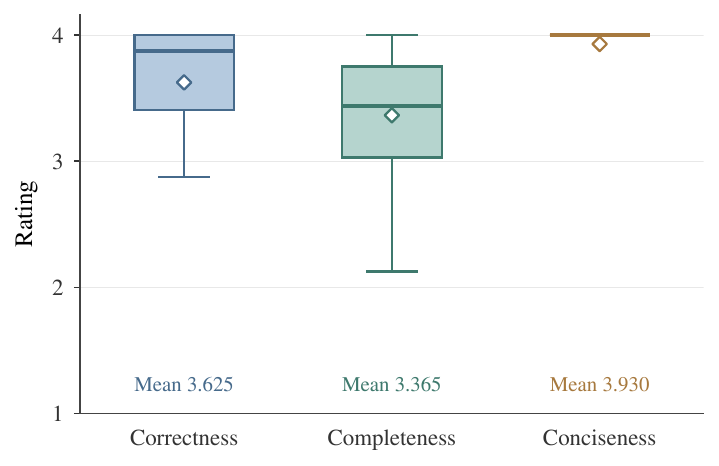}
    \caption{Human ratings of decision case quality.}\label{fig:decision-case-quality}
\end{wrapfigure}

As shown in Figure~\ref{fig:decision-case-quality}, the extracted decision cases
receive consistently high ratings. 
Conciseness receives the highest score, with a mean of 3.930, indicating that
the extracted cases generally avoid irrelevant or redundant information.
Correctness also achieves a high mean score of 3.625, suggesting that the
reconstructed content is largely faithful to the original review discussions
and patches. Completeness receives a slightly lower but still favorable mean
score of 3.365, indicating that most key information is successfully captured,
although some details from complex review discussions may still be omitted. 
Overall, the ratings are concentrated toward the upper end of the four-point
scale, providing evidence for the quality of the extracted decision cases.

\textit{LLM-Based Annotation.}
Manually labeling all 1,774 cases would be prohibitively costly. We
therefore employ three independent LLM annotators, GPT-6 Sol, DeepSeek-V4.1
Flash, and Claude Opus 5.5, to label each decision case for
FQ1 and FQ2, with majority vote determining the final label. We measure
inter-annotator agreement using Fleiss' $\kappa$~\cite{Landis1977TheMO}.

\subsection{Prevalence of Design Trade-offs (FQ1)}
\label{sec:prevalence}

We consider a decision case to involve a \emph{design trade-off} when it
presents multiple plausible alternatives whose resolution requires reasoning
about technical considerations or project-specific constraints. Then a review
thread is considered to involve a design trade-off if any extracted decision
case satisfies this definition.
Using this criterion, the three LLM annotators independently label the
sampled decision cases and achieve substantial agreement
(Fleiss' $\kappa=0.671$)~\cite{Landis1977TheMO}.

\begin{wrapfigure}{r}{0.62\textwidth}
    \centering
    \includegraphics[width=\linewidth]{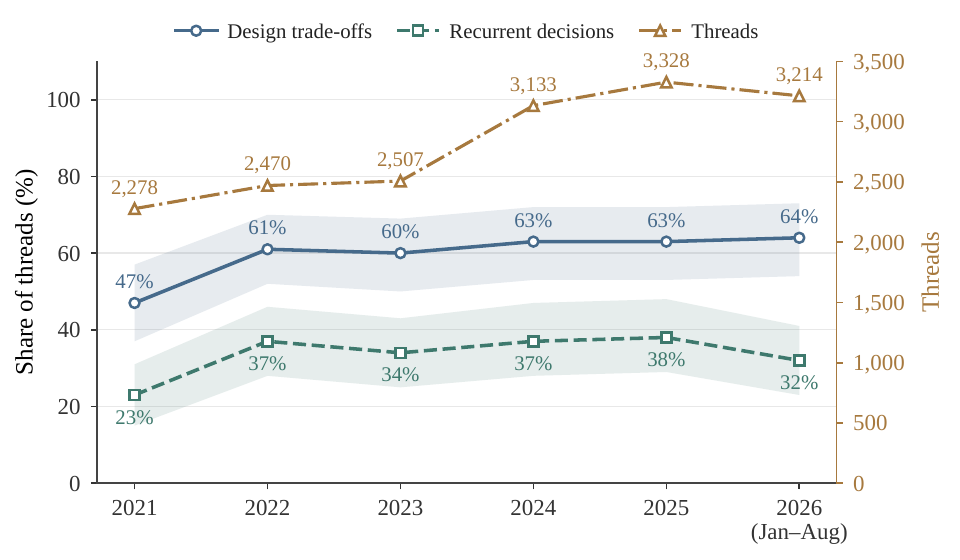}
    \caption{Yearly prevalence and recurrence.}
    \label{fig:prevalence_and_recurrence}
\end{wrapfigure}

\textit{Results.}
Overall, \textbf{60.29\%} of the sampled threads involve at least one design
trade-off. As shown in
Fig.~\ref{fig:prevalence_and_recurrence}, the yearly
prevalence increases from 47\% in 2021 to 61\% in 2022 and remains around
60--64\% thereafter. This pattern persists despite substantial variation in
yearly review volume, suggesting that design trade-offs are a recurring
feature of Linux kernel patch review rather than a phenomenon concentrated
in a particular period.

\textit{Example.}
In the \texttt{memfd} patch (Figure~\ref{fig:motivation}, left), an
allocation path could expose stale data in a huge page. Both
\texttt{folio\_zero\_range()} and \texttt{folio\_zero\_user()} clear the page,
but the former can clear it with \texttt{memset()} when
\texttt{CONFIG\_HIGHMEM} is disabled, whereas the latter processes constituent
pages with explicit rescheduling. Which one is preferable
depends on page size, architecture, and preemption configuration:
functional equivalence does not imply design equivalence.

\findingbox{Finding 1}{Design trade-offs are prevalent in Linux kernel
patch review: about 60\% of reviewed threads involve at least one decision
among multiple plausible alternatives, and this proportion remains stable over
time. Reviewing such changes requires reasoning beyond local correctness,
motivating the need for design-aware code review.}

\subsection{Recurrence of Design Decisions (FQ2)}
\label{sec:recurrence}

We consider a decision case \emph{recurrent} when an earlier case from a different review thread addresses the same underlying problem and can inform the later review. For each sampled
decision case, we retrieve the eight most semantically similar eligible
historical cases and ask the same annotators whether any candidate constitutes
such a precedent. A decision case is labeled recurrent if at least one such
historical precedent is identified. We then consider a review thread recurrent
if at least one of its extracted decision cases is recurrent.

\textit{Results.}
Overall, \textbf{33.91\%} of the sampled threads contain at least one recurrent
decision case, with substantial inter-annotator
agreement (Fleiss' $\kappa=0.605$). The yearly recurrence rate rises from 23\%
in 2021 and remains between 32\% and 38\% in subsequent years. 

Based on our manual inspection of some sampled threads, this LLM-based estimation is likely conservative. Annotators only examine the
eight most semantically similar candidates, so precedents that concern
different code or use different terminology, like the clearing discussion
relative to the \texttt{memfd} patch, may fall outside the candidate set and remain uncounted.

\textit{Example.}
An April 2025 patch series (Figure~\ref{fig:motivation}, right) proposed
clearing multiple pages together to improve throughput, and a reviewer
proposed bounded clearing batches that periodically allow rescheduling.
Although it concerns a different change, it exposes the same design question
as the later \texttt{memfd} patch: how aggressively memory can be cleared
without compromising scheduling responsiveness. The code change is new, but
the underlying design problem is not.

\findingbox{Finding 2}{Similar design decisions frequently recur
across independent review threads: roughly one third of reviewed threads
revisit a problem for which a relevant historical precedent exists, and this
estimate is likely conservative. Historical review discussions thus
offer a substantial opportunity for reusing design knowledge when reviewing
new patches.}

\subsection{Implications (FQ3)}
\label{sec:implications}

FQ1 motivates explicit support for reasoning about design trade-offs during
review, as well as a benchmark for this capability, and FQ2 shows that
historical discussions contain reusable precedents. These precedents,
however, are not directly reusable. Our study surfaces three challenges, each
motivating one stage of \system.

\noindent\textbf{C1: Recovering Coherent Design Decisions.}
A decision rarely appears in one place: in the clearing discussion, the
proposal, the problematic preemption condition, and the bounded alternative
appear in different replies, while the merged code records only the adopted
solution. Historical review knowledge therefore needs to be reconstructed into
decision-level units that connect the design problem, alternatives, trade-offs,
and final decision, rather than treated as raw emails.

\noindent\textbf{C2: Locating Relevant Precedents.}
Precedents may involve different files, features, and terminology, as in the
clearing discussion and the \texttt{memfd} patch. Our recurrence analysis
identifies similar problems from reviewer discussions, but at review time only
the new patch is available. Relevant precedents therefore need to be indexed
by the code situations that trigger them, rather than by surface-level
similarity in the discussion.

\noindent\textbf{C3: Applying Precedents to the Current Code.}
A historical resolution holds only if its conditions still hold: applying
the clearing precedent requires checking page sizes and preemption
configuration in the current code. Historical precedents therefore need to be
treated as hypotheses to be checked against the current code rather than
answers to be copied directly.

\findingbox{Finding 3}{\revise{Historical review discussions cannot be reused
as is. Effective reuse requires (C1) recovering decision-level knowledge from
fragmented discussions, (C2) locating precedents from the code situation of a
new patch rather than from surface text, and (C3) validating whether a
precedent still applies to the current code.}}

\section{\system: Knowledge-Augmented Code Review}
\label{sec:designlore}
We propose \system, a knowledge-augmented code review approach that recovers
reusable design knowledge from historical review discussions and leverages it
to support design-aware review of new patches.

\begin{figure*}[t]
  \centering
  \includegraphics[width=\textwidth]{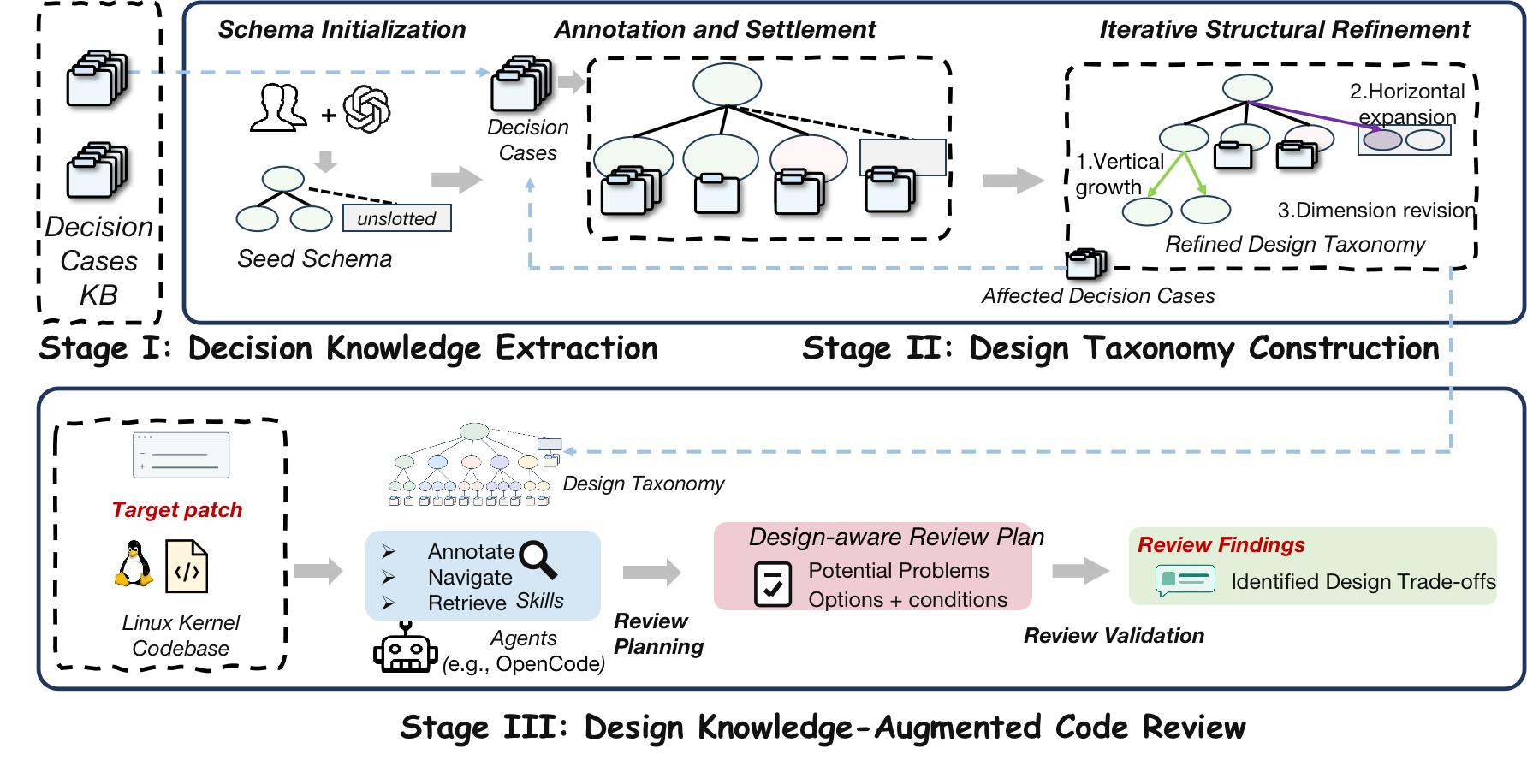}
  \caption{Overview of \system.}
  \Description{A three-stage pipeline divided into offline and online phases.
  Offline decision-knowledge extraction transforms historical review threads
  into source-linked decision cases, and taxonomy construction organizes the
  cases into a taxonomy-linked index. Online design-aware review routes a
  target patch through the taxonomy, retrieves eligible precedents, composes a
  review plan, and produces design findings after inspecting the target code.}
  \label{fig:pipeline}
\end{figure*}

As illustrated in Figure~\ref{fig:pipeline}, \system consists of two stages\revise{,
each addressing one of the challenges identified in the formative study}.
First, it recovers structured decision cases from historical review threads
through \revise{the} mining pipeline \revise{introduced earlier in our formative study}, \revise{which} progressively
transforms fine-grained atomic suggestions into coherent design decisions \revise{(C1)}.
Second, it organizes the recovered cases into a design taxonomy according to
their code-pattern triggers, forming a reusable project-level knowledge base
\revise{that can be navigated from the code of a new patch (C2)}.
At review time, \system uses the target
patch to navigate the knowledge base, retrieves relevant historical
precedents, synthesizes them into a decision-oriented review plan, and
validates the plan against the code under review \revise{(C3)}.

\subsection{Decision Knowledge Extraction}
\label{sec:designlore-extraction}

\revise{The first stage of \system addresses challenge C1 by applying the
mining pipeline of Section~\ref{sec:decision-extraction} to the entire
historical review corpus rather than to a sample of threads. Each thread is
thereby transformed into a decision-case tree, whose nodes are source-linked
decision cases and whose edges record how resolving one design question
exposed another. These cases form the raw material organized and reused by
the next two stages.}

\subsection{Design Taxonomy Construction}
\label{sec:taxonomy-construction}
The decision cases recovered in the previous stage preserve the reasoning
behind individual review discussions. A flat collection of such cases,
however, provides no stable routing structure for a new patch. \system
therefore organizes historical decision cases into a hierarchical design
taxonomy according to their \emph{Code-pattern trigger}. The taxonomy provides
a structured entry point for locating historical decisions that arise under
similar code circumstances.

Inspired by established taxonomy development methods~\cite{taxoadapt2025, atlas2026}, 
\system constructs the taxonomy using a round-based, top-down procedure that
combines a pre-established seed schema with data-driven adaptation.
Algorithm~\ref{alg:taxonomy-construction} summarizes the process.

\begin{algorithm}[t]
\caption{Round-based adaptive taxonomy construction}
\label{alg:taxonomy-construction}
\footnotesize

\noindent
\begin{minipage}[t]{0.47\linewidth}
\begin{algorithmic}[1]
\Require Decision cases $\mathcal{C}$; seed schema $\mathcal{T}_0$;
batch size $b$; admission threshold $a$; overload threshold $\tau$;
max rounds $K$; max depth $D$; max expansions $R$
\Ensure Design taxonomy $\mathcal{T}$

\Statex \Comment{Schema Initialization}
\State $\mathcal{T} \gets \Call{Initialize}{\mathcal{T}_0}$
\State $\mathcal{W}_1 \gets \mathcal{C}$

\For{round $r = 1,\ldots,K$}
    \Statex \Comment{Batch-Wise Annotation and Settlement}
    \For{each batch $B$ in $\Call{Batches}{\mathcal{W}_r,b}$}
        \State $S_B \gets \Call{Snapshot}{\mathcal{T}}$
        \ForAll{$c \in B$ \textbf{in parallel}}
            \State $(\pi_c,q_c) \gets \Call{Annotate}{c,S_B}$
        \EndFor
        \State $\mathcal{Z}_B \gets \{(c,\pi_c,q_c) \mid c \in B\}$
        \State $\mathcal{T} \gets \Call{Settle\_Batch}{\mathcal{T},\mathcal{Z}_B,a}$
    \EndFor
    \Statex \Comment{Iterative Structural Refinement}
    \State $\bar{S}_r \gets \Call{Snapshot}{\mathcal{T}}$
\algstore{taxo}
\end{algorithmic}
\end{minipage}
\hfill
\begin{minipage}[t]{0.51\linewidth}
\begin{algorithmic}[1]
\algrestore{taxo}
    \For{each node $n$ in $\Call{Postorder}{\bar{S}_r}$}
        \If{$n$ is a leaf \textbf{and} $|\mathcal{C}(n)|\!>\!\tau$ \textbf{and} $d(n)\!<\!D$}
            \State $\mathcal{T} \gets \Call{Grow\_Dimension}{\mathcal{T},n,R}$
        \ElsIf{$n$ is a routing node \textbf{and} $|U(n)|\!>\!\tau$}
            \If{$\Call{Expansion\_Count}{n} < R$}
                \State $\mathcal{T} \gets \Call{Expand\_Categories}{\mathcal{T},n}$
                \State $\Call{Increment\_Expansion\_Count}{n}$
            \Else
                \State $\mathcal{T} \gets \Call{Revise\_Dimension}{\mathcal{T},n}$
            \EndIf
        \EndIf
    \EndFor
    \State $\mathcal{W}_{r+1} \gets \Call{AffectedCases}{\bar{S}_r,\mathcal{T}}$
    \If{$\mathcal{W}_{r+1} = \varnothing$}
        \State \textbf{break}
    \EndIf
\EndFor
\State \Return $\mathcal{T}$
\end{algorithmic}
\end{minipage}
\end{algorithm}

\subsubsection{Schema Initialization.}
Upper-level taxonomy dimensions are intended to capture relatively stable and
fundamental properties of code changes, providing consistent routing semantics
for subsequent construction. To establish such a stable structure, \system
initializes the upper \(k\) levels through an iterative
induction-and-validation process. An LLM-based annotation agent induces, for
each level, a candidate routing dimension and its associated categories from a
sampled subset of decision cases. The resulting schema is then validated and
refined on disjoint batches until its dimensions and categories stabilize.
Finally, the stabilized upper-level schema is manually reviewed and confirmed
before being fixed as the seed taxonomy schema. Each routing node additionally
maintains an \emph{unslotted} bucket for cases that cannot be assigned to any
existing category.

\subsubsection{Batch-Wise Annotation and Settlement.}
Decision cases are processed in batches. At the beginning of each round,
\system freezes the current taxonomy as a snapshot so that all cases in the
batch are annotated against the same schema, reducing sensitivity to
processing order. An annotation agent routes each case top-down. At each node,
the agent selects an existing category according to the node's routing
dimension. If no category is applicable, the case is placed in the
corresponding \emph{unslotted} bucket and the agent proposes a candidate label
that remains within the same routing dimension.

To encourage consistent terminology, candidate labels are generated with
reference to the vocabulary already associated with the routing node. At the
end of each batch, \system normalizes semantically equivalent candidate labels
and aggregates their support. Candidate labels with sufficient recurring
evidence are admitted as new categories, after which the corresponding
unslotted cases are rerouted to the newly created nodes.

\subsubsection{Iterative Structural Refinement.}
After batch settlement, the resulting taxonomy may still be too coarse to
organize the accumulated cases effectively. \system therefore examines
overloaded nodes and applies three refinement operations. 
\emph{(1) Vertical growth} handles overloaded leaf
nodes by introducing a more specific routing dimension and its categories
beneath the leaf. \emph{(2) Horizontal expansion} handles routing nodes with
large unslotted buckets by introducing additional categories under the
existing dimension. \emph{(3) Dimension revision} is triggered when repeated
horizontal expansion fails to produce a meaningful partition; \system then
redefines the routing dimension of the affected adaptive subtree and
repartitions its cases. Expert-seeded dimensions are never revised.

A refinement agent performs these operations using two complementary forms of
context. \emph{Path context} contains the definitions and routing dimensions
along the path from the root to the current node, ensuring that a newly
introduced dimension is more specific than its ancestors. \emph{Peer context}
contains the definitions and routing dimensions of nodes at comparable depth,
helping maintain a consistent level of abstraction across the taxonomy. The
agent is additionally provided with decision cases from the node being
refined. After each structural change, all affected cases are re-annotated
against the updated local subtree. Refinement continues until no overloaded
node admits a valid structural improvement or a predefined refinement limit 
is reached, i.e., the maximum depth or retry count.

\subsection{Design Knowledge-Augmented Code Review}
\label{sec:code-review}

The previous offline stages construct a reusable knowledge base of historical
design decisions organized by the design taxonomy. In the online stage,
\system integrates this knowledge base into a coding agent through a set of
knowledge-access skills. Specifically, \system organizes the review process into two stages: \emph{knowledge-guided review planning} and \emph{code-grounded review validation}.

\subsubsection{Knowledge-Guided Review Planning}

Given a target patch and the repository snapshot under review, \system first
instructs the agent to inspect the changed code and its surrounding context,
identify salient code patterns, and formulate an initial characterization of
the affected functionality and implementation mechanism. The agent then uses
this characterization to explore the historical design knowledge base.

To support this exploration, \system provides three complementary
knowledge-access skills:

\begin{itemize}[leftmargin=12pt, topsep=3pt, itemsep=2pt, parsep=0pt, partopsep=0pt]
    \item \texttt{annotate}: maps code patterns identified in the target
    patch to candidate taxonomy nodes using the same classification procedure
    employed during taxonomy construction. This provides an initial entry
    point into the historical knowledge space;

    \item \texttt{navigate}: explores structurally and semantically related
    taxonomy nodes. The agent can navigate parent, child, and sibling nodes,
    and follow the \texttt{caused-by} relation to reach historically related
    situations, while inspecting the associated case statistics and
    representative decisions;

    \item \texttt{retrieve}: retrieves historical decision cases using
    BM25-based ranking. Retrieval can be performed globally over the complete
    knowledge base or locally within taxonomy nodes selected during
    navigation, allowing the agent to progressively narrow or expand the
    search scope.
\end{itemize}

Knowledge access proceeds iteratively through taxonomy navigation and case
retrieval. The agent first maps the target code to relevant taxonomy nodes,
then explores neighboring or causally related nodes and retrieves candidate
decision cases. Based on the retrieved evidence, it can refine the search
scope and continue exploring additional nodes as new relevant concerns emerge.
This iterative process progressively focuses the search on historical design
knowledge relevant to the current patch.

From the retrieved cases, the agent identifies recurring design concerns,
alternative solutions considered in prior reviews, their advantages and
disadvantages, and the conditions under which different solutions were
preferred. It synthesizes this information into a \emph{review plan}, which
contains a set of candidate design concerns for further examination. Each
plan item records the relevant code region, the potential design issue,
supporting historical precedents, possible alternative solutions, and the
conditions to be validated in the current repository. At this stage, each
plan item serves as a hypothesis for subsequent code-grounded validation.

\subsubsection{Code-Grounded Review Validation}
The second stage validates each planned concern against the current codebase.
For every review-plan item, the agent traces the affected code, relevant
control and data dependencies, and surrounding implementation context to
determine whether the historically observed design situation actually applies
to the target patch.

The agent further checks whether the assumptions and applicability conditions
associated with the historical cases hold in the current implementation,
such as lifecycle assumptions and interface constraints. 
Plan items that are not supported by the current code are discarded.
For the remaining items, the agent combines code evidence with the relevant
historical precedents to produce concrete \emph{review findings}. Each finding
identifies the underlying design concern, explains why it applies to the
current patch, and, when supported by both the historical evidence and the
current context, recommends a corresponding resolution with its rationale.
This validation stage prevents historically similar but
contextually inapplicable cases from being directly propagated into review
comments.

\section{Evaluation}
\label{sec:evaluation}

To comprehensively evaluate \system, we formulate the following research questions:

\begin{itemize}[leftmargin=12pt, topsep=3pt, itemsep=2pt, parsep=0pt, partopsep=0pt]
    \item \textbf{Overall Effectiveness (RQ1):}
    How effective is \system in improving design-aware code review compared
    with existing code-review approaches?

    \item \textbf{Generalizability (RQ2):}
    To what extent does \system retain its benefits across agent frameworks
    and backbone models?

    \item \textbf{Component Contribution (RQ3):}
    How much does each component of \system contribute to its design-aware
    review performance?

    \item \textbf{Reliability of Judgment (RQ4):}
    \revise{To what extent does the LLM-based evaluation align with human
    judgments?}
\end{itemize}

\subsection{\benchmark: A Design-Aware Code Review Benchmark}
\label{sec:benchmark}

We introduce \benchmark, a benchmark for evaluating design-aware code
review. Whereas existing code-review tasks primarily assess whether a
proposed patch contains correctness defects, \benchmark focuses on the
project-specific design trade-offs exposed by the patch.

\subsubsection{Task Definition}
We define \emph{design-aware code review} as the task of identifying
underlying design tensions or trade-offs exposed by a proposed code change.
Formally, given a patch series \(P\), that is, a set of related patches that
collectively implement a change, and the codebase snapshot \(C\) available at
review time, a reviewer outputs a set of findings \(\{f_i\}_{i=1}^{m}\), each
describing a design trade-off that may require further consideration during
review.

\subsubsection{Benchmark Construction}
We construct \benchmark through four phases: data collection, rule-based
filtering, LLM-based verification, and manual inspection. 

\noindent\textit{Step 1: Data Collection.}
We collect patch-review threads posted to the Linux kernel \texttt{mm} mailing
list from January 2025 to August 2026. We retain only threads associated with
patch submissions and containing substantive discussions among human reviewers,
resulting in 6,542 threads. For each retained thread, we apply the
\emph{Decision Case Extraction} pipeline (Section~\ref{sec:decision-extraction}), with GPT-5.6 Luna as the
underlying LLM, to reconstruct the corresponding
decision cases. These cases provide structured evidence for constructing
realistic design-aware code-review instances grounded in historical review
outcomes. Specifically, we use the recovered \emph{Problem} field as a
candidate design trade-off point.

\noindent\textit{Step 2: Rule-Based Filtering.}
We next apply deterministic filters to remove cases that cannot support a
reliable code-review instance. Specifically, we require that (i) the recovered
case contains at least two substantive alternatives, indicating that an actual
design choice was considered; (ii) the reviewed patch can be linked to the
corresponding kernel snapshot available at review time and can be applied to that
snapshot, ensuring that the benchmark instance has a reproducible code context;
and (iii) the patch series can be linked to its eventual mainline integration,
providing downstream evidence for validating the recorded resolution.

\noindent\textit{Step 3: LLM-Based Verification.}
We then perform semantic verification of the remaining candidates. Each
verifier is given the recovered case together with the original reviewed patch
and the corresponding final patch integrated into mainline. It independently
checks whether (i) the case captures a design trade-off rather than a
purely mechanical or straightforward correctness issue, with multiple
plausible alternatives under the available constraints; (ii) the recorded
final decision and rationale are sufficiently explicit and supported by the
review evidence; (iii) the eventual resolution differs meaningfully from the
design embodied in the original patch; and (iv) the recorded resolution is
reflected in the version ultimately integrated into mainline. We use GPT-6
Sol and DeepSeek-V4.1 Flash as independent verifier models and retain only
candidates accepted by both.

\noindent\textit{Step 4: Manual Inspection.}
Finally, two human annotators independently inspect the surviving candidates
and assess whether each case constitutes a valid design-aware code-review
instance. A candidate is retained when both annotators agree that the
design trade-off point, reference decision, and supporting rationale are
sufficiently clear and well grounded in the historical evidence. The two annotators achieve substantial agreement, with Cohen's $\kappa=0.637$.
This final curation yields \benchmark, consisting of 72 patch series and
79 design trade-offs.

\subsubsection{Evaluation Protocol}
\label{sec:evaluation_protocol}
We evaluate the findings produced by a code-review agent against the
reference design trade-offs in \benchmark.
Following the three-level detection scheme used in
Sashiko~\cite{sashiko2026}, each pair of a prediction and a reference is
independently evaluated by an LLM judge, which assigns one of three labels:
\emph{Exact}, \emph{Partial}, or \emph{Miss}.
\emph{Exact} indicates that the prediction identifies the same underlying
design trade-off as the reference; \emph{Partial} indicates that it captures
part of the trade-off but does not fully characterize the underlying tension;
and \emph{Miss} indicates that the target trade-off is not identified.
For quantitative evaluation, we map \emph{Exact}, \emph{Partial}, and
\emph{Miss} to graded scores of \(1\), \(0.5\), and \(0\), respectively,
and report graded precision, recall, and F1-score as the evaluation metrics.

Formally, let \(F\) denote the set of findings produced by an agent and
\(G\) the set of reference design trade-offs. For each pair of predictions and references\((f,g)\), let \(\ell(f,g)\) denote its assigned label and
\(s(f,g)=w(\ell(f,g))\) the corresponding graded score.
We compute graded precision, recall, and F1-score as: 
\[
P=
\frac{1}{|F|}
\sum_{f\in F}\max_{g\in G}s(f,g),
\qquad
R=
\frac{1}{|G|}
\sum_{g\in G}\max_{f\in F}s(f,g),
\qquad
F1=\frac{2PR}{P+R}.
\]

\subsection{Experimental Setup}
\label{sec:experimental-setup}

\subsubsection{Baselines}
We implement \system on top of OpenCode~\cite{anomaly2026opencode} and compare
it against six baselines. The first four represent
general-purpose or review-specialized coding agents. The remaining two are
retrieval-enhanced variants of OpenCode that directly access the raw historical
review-email corpus.

\begin{itemize}[leftmargin=12pt, topsep=3pt, itemsep=2pt, parsep=0pt, partopsep=0pt]
  \item Sashiko~\cite{sashiko2026} is an agentic reviewer developed by the
  Linux Foundation specifically for the Linux kernel, with kernel-specific
  prompts and a dedicated review protocol.
  \item OpenCodeReview~\cite{li2026opencodereview} is a review-specialized
  agent combining rule-guided dispatch, tool-grounded file inspection, and
  independent reflection.
  \item Codex~\cite{openai2026codex} is a general-purpose coding agent; we use
  its native review workflow.
  \item OpenCode~\cite{anomaly2026opencode} is an open-source coding agent; we
  use its native review workflow.
  \item OpenCode-RAG applies retrieval-augmented
  generation~\cite{lewis2020retrieval} over the raw historical review emails,
  providing semantically similar emails as additional context.
  \item OpenCode-Search lets OpenCode search the same raw-email corpus
  through keyword-based \texttt{grep} queries and inspect the retrieved emails
  as needed.

\end{itemize}

\subsubsection{Implementation Details}

\textit{Temporal Data Cut-off.}
To prevent temporal data leakage, the historical email corpus used by
\system and the retrieval-enhanced baselines contains only emails from
2021--2024, which strictly precede all benchmark patches (2025--2026).

\textit{Code Review.}
All baselines use GPT-5.6 Luna, controlling for the underlying model when
comparing review workflows. We also provide all agents with the same
design-aware code-review instruction, asking them to identify design 
tensions or trade-offs exposed by the target patch.
For OpenCode-RAG, we encode historical emails using
\texttt{sentence-transformers/all-MiniLM-L6-v2} and retrieve the top-8 most
similar emails for each target patch.

\textit{LLM Judges.}
We use three SOTA LLMs from distinct model families, namely GPT-6 Sol,
DeepSeek-V4.1 Flash, and Claude Opus 5.5, as independent judges.
Each judge evaluates agent findings using the criteria defined in
Section~\ref{sec:benchmark}. We average the resulting \revise{graded} scores
across judges to obtain the final scores, reducing dependence on
any single evaluator.

\subsection{Results and Analysis}

\subsubsection{RQ1: Overall Effectiveness}
\label{sec:rq1}

We evaluate \system against six baselines on design trade-off detection.
Table~\ref{tab:rq1-main-results} reports the results.
Overall, \system achieves the highest precision, recall, and F1 among all
evaluated methods, with an F1 of 0.259.

\begin{wraptable}{r}{0.55\columnwidth}
  \centering
  \vspace{-3pt}
  \caption{Overall Results on \benchmark.}
  \label{tab:rq1-main-results}
  \small
  \resizebox{\linewidth}{!}{
    \begin{tabular}{*{5}{c}}
      \toprule
        \multirow{2}{*}{\textbf{Method}} & \multicolumn{3}{c}{\textbf{Performance}} & \textbf{Cost} \\
      \cmidrule(lr){2-4} \cmidrule(lr){5-5}
        & Precision & Recall & F1-Score & \$/item \\
        \midrule
        OpenCode & 0.136 & 0.236 & 0.173 & 0.033 \\
        Codex & 0.173 & 0.272 & 0.211 & 0.021\\
        \midrule
        OpenCodeReview & 0.087 & 0.274 & 0.132 & 0.119 \\
        Sashiko & 0.151 & 0.363 & 0.213 & 0.101 \\
        \midrule
        OpenCode-Search & 0.155 & 0.226 & 0.184 & 0.102 \\
        OpenCode-RAG & 0.183 & 0.249 & 0.211 & 0.058 \\
        \midrule
        \system & \textbf{0.189} & \textbf{0.411} & \textbf{0.259} & 0.050 \\
    \bottomrule
    \end{tabular}
  }
  \vspace{-5pt}
\end{wraptable}

\textit{Comparison with general-purpose agents.}
\system builds on OpenCode, making their comparison a direct assessment of
the benefit introduced by our knowledge-augmented review workflow.
\system improves F1 from 0.173 to 0.259, a 49.7\% relative gain, while
improving both precision and recall.
It also outperforms Codex, the stronger general-purpose agent baseline,
whose F1 reaches 0.211.
These results show that \system benefits substantially
from augmenting its underlying agent with structured design
knowledge.

\textit{Comparison with specialized reviewers.}
The specialized review systems still show limited capability in
design-aware code review. Of the two, Sashiko is the stronger baseline,
with an F1 of 0.213, while OpenCodeReview achieves only 0.132.
This suggests that existing review-focused agents still struggle to
reliably identify the design trade-offs captured by our benchmark.
In contrast, \system outperforms both systems. Compared with Sashiko,
it achieves an F1 of 0.259, corresponding to a 21.6\% relative gain,
while also attaining higher precision and recall.

\textit{Comparison with historical-email retrieval.}
To isolate the benefit of \system's knowledge workflow, we further compare
against retrieval-augmented baselines that provide direct access to historical
review discussions. Compared with the underlying OpenCode agent, the stronger
retrieval baseline, OpenCode-RAG, improves F1 from 0.173 to 0.211, showing
that historical review information is useful for design-aware review.
\system further increases F1 to 0.259, corresponding to a 22.7\% relative
gain over OpenCode-RAG. The improvement is particularly pronounced in recall,
indicating that \system more effectively exploits historical design knowledge
to identify trade-offs in target patches. 
These results suggest that the gain comes not merely from access to raw
historical emails, but from structuring and organizing historical knowledge
for effective retrieval and reasoning.

\textit{Cost.}
The improved effectiveness of \system comes with a moderate cost overhead
over its underlying OpenCode agent, increasing the average review cost from
\$0.033 to \$0.050 per patch. Nevertheless, \system remains less expensive
than the stronger retrieval baseline OpenCode-RAG (\$0.058) and requires
roughly half the cost of Sashiko (\$0.101). 
The one-time offline cost of
constructing the knowledge base is discussed in Section~\ref{sec:discussion}.

\rqsummary{RQ1}{
Existing code-review agents still show limited capability in design-aware
review, whereas \system achieves the best design trade-off detection
performance among all evaluated methods, with an F1 of 0.259, 
demonstrating its effectiveness for design-aware review.
}

\subsubsection{RQ2: Generalizability}
\label{sec:rq2}

We examine the generalizability of \system along two dimensions:
the underlying agent system and the backbone LLM. First, we integrate
\system with multiple coding agents and compare each augmented configuration
against its corresponding vanilla agent. We then vary the backbone LLM and
compare \system with Sashiko, the strongest baseline in
RQ1, to examine whether the observed advantage persists across backbones.

\begin{wrapfigure}{r}{0.55\textwidth}
  \centering
  \includegraphics[width=\linewidth]{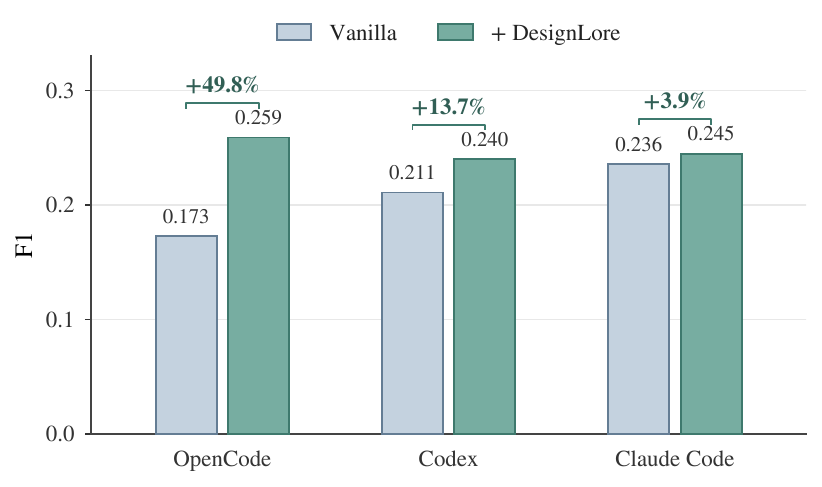}
  \caption{F1 improvements from \system across different underlying agents.}
  \label{fig:agent-generalizability}
\end{wrapfigure}

\textit{Generalizability across agent systems.}
We first evaluate whether the benefit of \system extends beyond its default
OpenCode configuration. 
In addition to the default OpenCode configuration with GPT-5.6 Luna,
we integrate \system with Codex using the same GPT-5.6 Luna backbone,
and with Claude Code~\cite{anthropic2026claudecode} using Claude Sonnet 5. 
Then we compare each augmented agent against its vanilla
counterpart. Figure~\ref{fig:agent-generalizability} reports the resulting F1
scores and relative gains. 
\system improves F1 across all three agent configurations, demonstrating 
its effectiveness.
The largest gain is observed on OpenCode, while an improvement is also
achieved on Claude Code with the Claude Sonnet 5 backbone, yielding a 3.9\% relative F1 gain. Overall, the consistent gains across these configurations suggest that \system can improve
design-aware code review without relying on behaviors specific to a single
harness.

\begin{table*}[t]
  \caption{Performance of \system across different backbone models.}
  \label{tab:model-generalizability}
  \centering
  \small
  \resizebox{\linewidth}{!}{
  \begin{tabular}{lcccccccccccc}
    \toprule
    \multirow{2}{*}{Method}
    & \multicolumn{3}{c}{GPT-5.6 Luna}
    & \multicolumn{3}{c}{GPT-5.6 Terra}
    & \multicolumn{3}{c}{Claude Haiku 4.5}
    & \multicolumn{3}{c}{DeepSeek-V4.1 Flash} \\
    \cmidrule(lr){2-4}
    \cmidrule(lr){5-7}
    \cmidrule(lr){8-10}
    \cmidrule(lr){11-13}
    & Precision & Recall & F1
    & Precision & Recall & F1
    & Precision & Recall & F1
    & Precision & Recall & F1 \\
    \midrule
    Sashiko
    & 0.151 & 0.363 & 0.213
    & 0.158 & 0.357 & 0.218
    & 0.077 & 0.228 & 0.115
    & 0.170 & 0.509 & 0.255 \\
    \system
    & 0.189 & 0.411 & 0.259
    & 0.202 & 0.409 & 0.271
    & 0.109 & 0.304 & 0.160
    & 0.192 & 0.649 & 0.297 \\
    \bottomrule
  \end{tabular}
  }
\end{table*}

\textit{Generalizability across backbone models.}
We next examine whether \system retains its advantage when the backbone LLM
changes. In addition to the default GPT-5.6 Luna configuration, we run both \system and Sashiko with three additional backbone models:
GPT-5.6 Terra, Claude Haiku 4.5, and DeepSeek-V4.1 Flash, while keeping the
evaluation protocol unchanged. Table~\ref{tab:model-generalizability} reports
the results.
\system consistently outperforms Sashiko across all three metrics with all four backbone models. The relative F1 gains range from 16.5\% to 39.1\%,
showing that the advantage of \system persists across models from different
families and with substantially different absolute performance levels. 

\rqsummary{RQ2}{
\system generalizes beyond its default configuration, yielding consistent
performance gains across multiple agent systems and outperforming the
strongest specialized-review baseline across all evaluated backbone models.
}

\subsubsection{RQ3: Component Contribution}
\label{sec:rq4}

\system consists of three sequential stages:
\emph{Decision-Knowledge Extraction} (Stage 1),
\emph{Design Taxonomy Construction} (Stage 2), and
\emph{Design Knowledge-Augmented Code Review} (Stage 3).
Since each later stage builds on the artifacts produced by the preceding
stages, we conduct a cumulative ablation study to quantify the incremental
contribution of each stage.
Starting from the full \system pipeline, we progressively remove the three
stages as follows: \emph{w/o Stage~3} removes the planning and
validation workflow, letting OpenCode directly use the taxonomy and retrieval
tools; \emph{w/o Stages~2--3} further removes the taxonomy, letting OpenCode
directly retrieve decision cases; and \emph{w/o Stages~1--3} reduces the
system to the original OpenCode baseline.

\begin{wraptable}{r}{0.49\columnwidth}
  \caption{Ablation results of \system.}
  \label{tab:ablation-study}
  \centering
  \resizebox{\linewidth}{!}{
  \begin{tabular}{lrrr}
    \toprule
    Method & Precision & Recall & F1-Score \\
    \midrule
    \system
      & \textbf{0.189} & \textbf{0.411} & \textbf{0.259} \\
    \quad \emph{w/o Stage 3}
      & 0.179 & 0.207 & 0.192 \\
    \quad \emph{w/o Stages 2--3}
      & 0.159 & 0.217 & 0.184 \\
    \quad \emph{w/o Stages 1--3}
      & 0.136 & 0.236 & 0.173 \\
    \bottomrule
  \end{tabular}
  }
\end{wraptable}

Table~\ref{tab:ablation-study} reports the cumulative ablation results.
Removing stages from \system leads to a monotonic decrease in F1, from
0.259 for the full system to 0.173 when all three stages are removed,
indicating that each stage contributes to the overall design-aware review
performance.

Among the three stages, Stage~3, the design knowledge-augmented review
workflow, provides the largest gain, increasing F1 from 0.192 to 0.259.
This result indicates that the planning and validation workflow helps the
agent more effectively exploit the available design knowledge and identify
additional design trade-offs. 
Stages~1 and~2 also provide consistent improvements. Making the extracted
decision knowledge available to OpenCode increases F1 from 0.173 to 0.184,
while further organizing this knowledge with the design taxonomy raises F1
to 0.192. These gains suggest that historical decision knowledge and its
structured organization help the agent focus on more relevant design
concerns.

\rqsummary{RQ3}{
The cumulative ablation study shows that each stage of \system contributes
to its overall design-aware review performance.
}

\subsubsection{RQ4: Reliability of Judgment}
\label{sec:rq5}

\revise{Our evaluation relies on LLM judges to assess agent-generated review
findings.} \revise{We therefore} assess whether the LLM-based evaluation used in
our experiments aligns with human judgments. We randomly sample 100 pairs of agent-generated
review findings and reference design trade-offs, together with the averaged scores assigned by the three LLM judges.
Following prior work~\cite{DBLP:journals/pacmse/WangGGFCX25}, we use two
independent human evaluators to construct the human reference scores.
Each evaluator assesses every sampled pair using our three-level rubric,
namely \emph{Exact}, \emph{Partial}, and \emph{Miss}. 
We first examine agreement between the two human evaluators using weighted
Cohen's \(\kappa\), and then assess the alignment between the averaged human
scores and the averaged LLM judge scores using Pearson's \(R\), Spearman's
\(\rho\), and Kendall's \(\tau_b\).

The two human evaluators achieve high agreement
($\kappa=0.842$), indicating that the human reference judgments are stable.
The ensemble LLM scores also align closely with human evaluation, achieving
Pearson's $R=0.945$, Spearman's $\rho=0.944$, and Kendall's
$\tau_b=0.890$. The consistently high agreement and correlations provide
evidence that the LLM-based evaluation reliably reflects human judgments on
design trade-off identification.

\rqsummary{RQ4}{
\revise{The} LLM-based evaluation shows strong alignment with human judgments.
}

\section{Discussion}
\label{sec:discussion}

\textbf{Toward Design-Aware Code Review.}
Our results suggest that recognizing design trade-offs and reasoning about
appropriate alternatives remain challenging for current review agents.
Unlike defect detection, a design concern may involve multiple viable
solutions whose suitability depends on project-specific constraints and
rationale. This calls for review systems that go beyond identifying problematic
code to explicitly reason about design decision points, competing alternatives,
and their trade-offs. Historical design knowledge provides a promising source
of such context, enabling review agents to relate a new patch to how similar
design choices were previously discussed.

\textbf{Toward an Evolving Project Memory.}
\system currently reconstructs decision cases offline from historical review
discussions. A natural extension is to incrementally incorporate newly
completed reviews, allowing these cases to form a continuously evolving
project-specific design memory. Such reuse, however, should remain
context-sensitive: a historical decision reflects the constraints and
assumptions under which it was made and should not be treated as a prescriptive
rule. Future systems should therefore retrieve relevant precedents while also
reasoning about whether their applicability conditions still hold in the
current code and project context.

\textbf{Offline Construction Cost.}
Building the project-specific design knowledge base incurs an upfront cost
because \system processes historical review discussions through multiple
LLM-based stages. In our study, extracting decision cases from 8,382 review
threads costs \$307.9, while constructing the design taxonomy incurs an
additional \$16.7. This yields a total offline construction cost of \$324.6. 
This cost is paid only during knowledge-base construction and can be amortized
across subsequent reviews. Once the historical archive has been processed,
newly arriving review discussions would only require incremental extraction and
taxonomy updates. The practical trade-off is therefore between the one-time
cost of constructing and maintaining project-specific design knowledge and
its repeated reuse during future code reviews.
\section{Related Work}
\label{sec:related}

\subsection{Design Knowledge Extraction}

Software development continuously involves design trade-offs and decisions, whose underlying knowledge is important for the long-term evolution of software systems. Prior work has therefore explored explicitly representing and preserving such knowledge to mitigate its loss over time. Jansen and Bosch modeled software architecture as a composition of explicit design decisions~\cite{jansen2005software}, while subsequent studies extended this view to capture alternatives, rationale, assumptions, context, and decision history, and further regarded architectural decisions as reusable knowledge assets~\cite{kruchten2006building,zimmermann2011architectural}. These studies establish the value and key elements of design knowledge, but largely assume that such knowledge is explicitly documented and maintained during development.

Another line of research focuses on recovering design knowledge from development artifacts, motivated by the observation that important design decisions are often discussed and refined through informal communication rather than explicitly documented~\cite{cubranic2003hipikat,brunet2014developers,sutherland2009peer,bi2021architecture,alsafwan2022rationale,bhat2017automatic,elzanaty2018design,DBLP:conf/icse/SharmaSS21}. More recently, researchers have leveraged LLMs to recover increasingly structured forms of design knowledge from such artifacts. Dhaouadi et al.~\cite{DBLP:conf/kbse/DhaouadiOF22,dhaouadi2025automated} constructed rationale and decision graphs from development histories, while DRMiner~\cite{DBLP:conf/kbse/ZhaoY0LY024,DBLP:conf/kbse/ZhaoY0LY024a} combines LLMs with heuristic signals to recover proposed solutions together with supporting and opposing arguments from issue discussions. Yu et al.~\cite{DBLP:journals/corr/abs-2604-05955} extracted project-specific design constraints from code review discussions arising during issue resolution. Other recent studies further explore LLMs for generating or recovering architectural rationale from textual artifacts~\cite{DBLP:journals/corr/abs-2504-20781}. Collectively, these studies show that development artifacts provide a rich source for recovering otherwise implicit design knowledge.

\subsection{Automated Code Review}

Early learning-based approaches formulate code review as prediction over code changes and review comments. Tufano et al.~\cite{tufano2021automating} studied pre-review transformation and comment-guided refinement, while CodeReviewer~\cite{li2022codereviewer} introduced review-specific pretraining for change-quality estimation, comment generation, and code refinement. Retrieval-based approaches such as CommentFinder~\cite{hong2022commentfinder} and LAURA~\cite{zhang2025laura} further exploit historical reviews to improve comment generation. More recent systems increasingly combine repository context, tool use, and multi-stage verification. CodeAgent~\cite{tang2024codeagent}, RovoDev~\cite{tantithamthavorn2026rovodev}, OpenCodeReview~\cite{li2026opencodereview}, and Sashiko~\cite{sashiko2026} organize review around repository exploration, specialized guidance, and validation of generated findings. Historical discussions can also provide reusable review knowledge: FLINT~\cite{DBLP:journals/corr/abs-2603-24825} distills Linux review discussions into validation rules and retrieves relevant rules and source context for new patches.

Existing benchmarks evaluate complementary aspects of automated code review. CodeReviewer~\cite{li2022codereviewer} and CodeReviewQA~\cite{lin2025codereviewqa} assess review generation, refinement, and comprehension, while DeepCRCEval~\cite{lu2025deepcrceval}, CRScore~\cite{naik2025crscore}, and HalluJudge~\cite{tantithamthavorn2026hallujudge} focus on review quality and grounding. More recent benchmarks move toward realistic agent-level evaluation with broader repository context: SWR-Bench~\cite{zeng2026swrbench} evaluates issue coverage over complete pull requests, ContextCRBench~\cite{hu2026contextcrbench} incorporates issue and program context, CR-Bench~\cite{DBLP:journals/corr/abs-2603-11078} evaluates defect detection together with review usefulness and noise, and c-CRAB~\cite{DBLP:journals/corr/abs-2603-23448} evaluates code-review agents on real pull requests by converting human-identified review concerns into executable tests.

Our work focuses on implicit design trade-offs, requiring the reviewer to \revise{recognize the design trade-off exposed by a patch and reason about it under project-specific constraints}. While FLINT distills historical Linux review knowledge into validation rules, \system reconstructs decision cases that preserve the problem, competing alternatives, and their underlying trade-offs. Accordingly, \benchmark focuses on whether an agent can identify such \revise{design trade-offs from a proposed patch}.
\section{Threats to Validity}
\label{sec:threats}

\textbf{Internal Validity.}
A key threat is the reliance on LLMs in multiple stages. We mitigate it by
using multiple independent LLMs for verification and evaluation, and validate
both the extracted decision cases (Section~\ref{sec:annotation}) and the
LLM-based evaluation (RQ4) through human assessment. Second, the candidate
trade-offs in \benchmark are recovered by the same extraction pipeline that
builds the knowledge base of \system, which may favor its representation. To
reduce this risk, every reference is verified against the original and
mainline patches by two LLM verifiers and two human annotators, and the
knowledge base contains only discussions that precede all benchmark patches.
Third, the backbone LLMs may have been exposed to the benchmark discussions
during pre-training; since all methods share the same backbone, this exposure
affects them equally.

\textbf{External Validity.}
Our study focuses on the Linux kernel memory-management subsystem
(\texttt{mm}), a representative and mature software system with long-running
patch-review discussions and traceable patch evolution. The prevalence of
design trade-offs and the effectiveness of \system may differ across other 
repositories. However, we believe that the underlying principle of recovering and
reusing historical design knowledge can generalize to other projects with
rich developer discussions, including pull-request-based development
platforms. Further validation on additional repositories is
left for future work. In addition, although the main experiments use
GPT-5.6 Luna as the backbone model, we evaluate \system with additional LLMs,
such as \revise{DeepSeek-V4.1 Flash}, and observe consistent improvements, suggesting that
the approach is not tied to a single model.

\textbf{Construct Validity.}
The reference design trade-offs in \benchmark are derived from documented
developer discussions and subsequent patch evolution. Therefore, they may not
capture implicit design considerations that were never explicitly discussed
or recorded. Consequently, our benchmark measures observable design reasoning
rather than the complete set of design considerations involved in software
maintenance. Nevertheless, it provides a practical and reproducible
approximation of design-aware code review beyond conventional correctness
checking.
\section{Conclusion}
\label{sec:conclusion}

A formative study of Linux kernel \texttt{mm} review discussions shows that
design trade-offs are prevalent and that similar problems frequently recur
across independent patch series. Building on these findings, we develop
\system, which recovers, organizes, and reuses historical decision cases, and
construct \benchmark for systematic evaluation. Existing coding agents show
limited capability in design-aware review, whereas \system achieves the
strongest performance and generalizes across agent systems and backbone
models. These results underscore the importance of design-aware review and
demonstrate that explicitly reco vering and structuring historical design
knowledge can improve automated code review beyond local correctness checking.

\bibliographystyle{ACM-Reference-Format}
\bibliography{references}

\end{document}